\documentclass{article}
\usepackage[english]{babel}
\usepackage[utf8]{inputenc}
\usepackage{geometry}
\usepackage{caption}
\usepackage{subcaption}
\usepackage{color, soul}
\usepackage{amsmath}
\usepackage{amssymb}
\usepackage{authblk}
\usepackage{multirow}
\usepackage{array}
\usepackage{booktabs}
\title{Absorption and birefringence study for reduced optical losses in diamond with high NV concentration}
\author[1]{T. Luo}
\author[1]{F. A. Hahl}
\author[1]{J. Langer}
\author[1]{V. Cimalla}
\author[1]{L. Lindner}
\author[1]{X. Vidal}
\author[1]{M. Haertelt}

\author[2]{R.~Blinder}
\author[3]{S. Onoda}
\author[3]{T. Ohshima}
\author[1]{J. Jeske}
\affil[1]{Fraunhofer Institute for Applied Solid State Physics IAF, 79108 Freiburg, Germany}
\affil[2]{Ulm University, D-89081 Ulm, Germany}
\affil[3]{National Institutes for Quantum Science and Technology (QST), 1233 Watanuki, Takasaki, Gunma 370-1292, Japan}

\usepackage[numbers]{natbib}
\usepackage{graphicx}
\usepackage{hyperref}
\hypersetup{colorlinks,allcolors=black}

\usepackage[normalem]{ulem} %need for strikethrough
\usepackage{xcolor}
\definecolor{orange}{rgb}{0.7,0.2,0}
\definecolor{darkgreen}{rgb}{0,0.5,0}

\begin{document}

\maketitle
\pdfminorversion=5

Accepted by Philosophical Transactions A (DOI: 10.1098/rsta.2022.0314).
\begin{abstract}
The use of diamond color centers such as the nitrogen-vacancy (NV) center is increasingly enabling quantum sensing and computing applications.
Novel concepts like cavity coupling and readout, laser threshold magnetometry and multi-pass geometries allow significantly improved sensitivity and performance via increased signals and strong light fields. 
Enabling material properties for these techniques and their further improvements are low optical material losses via optical absorption of signal light and low birefringence.
Here we study systematically the behavior of absorption around 700~nm and birefringence with increasing nitrogen- and NV-doping, as well as their behavior during NV creation via diamond growth, electron beam irradiation and annealing treatments. 
Absorption correlates with increased nitrogen-doping yet substitutional nitrogen does not seem to be the direct absorber. 
Birefringence reduces with increasing nitrogen doping. 
We identify multiple crystal defect concentrations via absorption spectroscopy and their changes during the material processing steps and thus identify potential causes of absorption and birefringence as well as strategies to fabricate CVD diamonds with high NV density yet low absorption and low birefringence.
\end{abstract}

\section{Introduction}
Negatively charged nitrogen-vacancy (NV) centers in diamonds have been extensively studied for quantum sensing applications.
Since their spin property can be optically read out at room temperature, NV centers have become a leading experimental quantum system and are widely used to detect magnetic field~\cite{balasubramanian2008nanoscale,maze2008nanoscale,degen2008scanning,acosta2009diamonds}, electric field~\cite{dolde2011electric,dolde2014nanoscale}, temperature~\cite{acosta2010temperature,kucsko2013nanometre,neumann2013high}, pressure~\cite{doherty2014electronic} and strain~\cite{ovartchaiyapong2014dynamic,teissier2014strain} and are a promising system to realise advanced sensing concepts such as field fluctuation sensing and identification of quantum coherent systems in the environment \cite{Coledecoherencemicroscopy,Jeske2012,luan2015decoherence}.
NV-ensembles in bulk diamonds can provide significantly stronger signals than individual NV centers for applications that require good sensitivity, which can be further enhanced by the multi-pass readout~\cite{zhou2014quantum,clevenson2014enhanced}, optical cavity coupling~\cite{faraon2012coupling,fehler2019efficient,dumeige2019infrared,ruf2021resonant} and laser cavity sensing~\cite{jeske2016laser, dumeige2019infrared, Nair2020, Nair2021,hahl2022magnetic}. 
Such techniques can significantly increase the readout signal and/or contrast by improved collection from the NV centers, increased coupling to a collection mode or stimulated emission. For their performance, a material parameter becomes particularly relevant: low optical losses of the signal light in the diamond material. Optical losses in diamonds are typically small and thus play a minor role in normal NV fluorescence collection. For the above-mentioned techniques, however, the level of optical losses can limit or significantly enhance sensitivity advantages. 
These techniques further benefit from high NV concentrations for a strong signal~\cite{barry2020sensitivity}, while the crystal quality also plays an important role as it enables a reduced optical loss in the material.

Diamond absorption of the light signal is the main source that introduces optical loss~\cite{hahl2022magnetic}.
Specifically, absorption at around 700~nm is the most relevant regime, as it contains most of the NV fluorescence.
As an example, stimulated emission of NV centers~\cite{jeske2017stimulated} and laser-threshold magnetometry (LTM), i.e.~the idea to use NV centers as a gain medium~\cite{jeske2016laser, Nair2020, Nair2021, savvin2021nv, hahl2022magnetic, fraczek2017laser} are ways to significantly enhance sensitivity.
For this, a very low absorption at $\sim$700~nm is needed to achieve net optical gain and lasing~\cite{hahl2022magnetic}.
Understanding the source of the absorption in this regime plays a crucial role to optimize the material quality thus improve the performance of the sensing method.

Previous studies identified many absorbers in diamond, which can give us hints of sources of the absorption at 700~nm, here we list four point defects as potential candidates: 
The first one is the H2 center (NVN$^-$), which has a zero-phonon-line (ZPL) at 986.3~nm and a very broad absorption side band from around 600~nm to the ZPL~\cite{dobrinets2016hpht}.
It often appears when creating NV centers, and it is also formed from the single substitutional nitrogen atoms (denoted as P1 center, N$^0_s$ or C-center). 
% When heating up the diamond, H2 centers are stable until 2100$^{\circ}$C~\cite{collins2005high}, while NV centers becomes unstable above 1000$^{\circ}$C~\cite{orwa2011engineering,luhmann2018screening}. 
% This brings difficulties to prevent the formation of H2 centers while pursuing high NV concentrations.
The second one is a band centered at 730~nm, which is always present together with 520, 552 and 840~nm bands, but not correlated perfectly in intensity~\cite{dobrinets2016hpht}, and its origin has not been well defined.
% This band has been observed in mainly high-nitrogen, high-hydrogen (HPHT) but not chemical vapor deposition (CVD) diamonds.
The third candidate is a broad band centered at 710~nm, which may be associated with the vibronic side-band of the nickel-related center at 794~nm~\cite{shigley1993gemological}.
The last one is then the GR1 band caused by neutral single vacancies (V$^0$), which has a ZPL at 741~nm and a broad
feature from around 500 to 750~nm~\cite{zaitsev2018defect}.
Among these candidates, the first one is commonly observed in chemical vapor deposition (CVD) diamonds, while the second and third appear most likely in high-pressure, high-temperature (HPHT) diamonds. 
The last one then appears mostly in irradiated diamonds. 
Apart from the point defect, non-diamond inclusions can also play a role as they contribute to a high absorption in the whole UV-Visible range, also resulting in a grayish coloration of the diamond~\cite{zaitsev2020nitrogen}.

Another distinct source of optical loss in diamond is the birefringence.
For cavity-based NV-magnetometry, optical modes suffer from birefringence in the material.
Moreover, polarisation-selective addressing and readout of NV centers are also only possible without birefringence as well as polarisation-selective elements, such as Brewster's angle diamond interfaces and optical elements.
Therefore, when pursuing high NV concentrations for improved sensitivity, the influence of the nitrogen content on birefringence also needs to be considered.
Diamond is generally considered to be an isotropic crystal without or with weak birefringence. 
However, extended defects in diamonds such as dislocations and stacking faults generate strain field that distort the diamond lattice~\cite{pinto2009theory,tallaire2017thick}.
This leads to localized birefringence patterns with randomly varying slow axes~\cite{friel2010development}, which appear inhomogeneously in the crystal. 
Additionally, applied stress can also make the diamond structure anisotropic thus resulting in birefringence.

Diamonds with high NV concentrations, low absorption and low birefringence lead to a strong signal and reduced optical loss, thus benefiting applications with large sensing volumes.
In this work, we investigate the link between nitrogen-doping, NV creation and diamond absorption/birefringence, discuss possible causes of absorption and birefringence, and suggest potential approaches to reduce them to obtain an optimized diamond.

\section{Methods}
\subsection{Sample processing}
We have grown over fifty (100) oriented CVD diamonds with varying parameters for both the absorption and birefringence study.
The growth of all CVD samples was run in an ellipsoidal-shaped CVD reactor with a 2.45 GHz microwave frequency and equipped with a 6 kW microwave generator~\cite{funer1998novel}.
Apart from CVD samples, we also investigated HPHT Ib diamonds (Element Six and Sumitomo) for the birefringence study.
All samples are in the form of bulk diamond plates with a geometry of 3$\times$3~mm$^2$ or 4$\times$4~mm$^2$, the thickness is varied from 200-1400~$\mu$m.
They have been cut from the substrate and polished on both surfaces with a roughness Ra<0.5~nm before the optical measurement.
The samples are investigated after growth, and partially after irradiation and annealing.
We have conducted electron-beam irradiation with 1~MeV and 2~MeV electron energies and different fluences, the subsequent annealing step was performed at 1000$^{\circ}$C for 2~h.

\subsection{Optical measurements}
The absorption of the sample was measured with a UV-Vis spectrometer (PerkinElmer Lambda 950) at room temperature.
The absorption coefficient was obtained from the transmittance $T$, which was measured by the spectrometer.
% The transmittance $T$ was measured by the spectrometer, then used to deduce the absorption coefficient.
For a precise study of absorption at $\sim$700~nm, we measured the sample with an integrating sphere at the wavelength of 680-760~nm to gather all transmitted light in this regime.
The absorption coefficient $A_{coef}$ in this case is given by~\cite{Hahl2022LTM}:
\begin{equation}
    \label{eq:Acoeff}
    A_{coef}=-\frac{\log_{10}(\sqrt{4T^2+(1-2R_{t}+R_{t}^2-T^2)^2}-1+2R_{t}-R_{t}^2+T^2)}{2dT}
\end{equation}
where $d$ is the sample thickness and $R_{t}\approx29.13\%$ is the theoretical value of the reflectance 
based on the Fresnel equation for normal incidence on a diamond with refractive index $n=2.4$.
For both the absorption and birefringence calculation (discussed below), $d$ is given by the average thickness of the diamond plate.

For the spectral study on the other hand, we used the standard detector of the spectrometer without the integrating sphere to measure the transmittance at 200-800~nm (since the functional wavelength of the material in the integrating sphere is only above 400~nm, the sphere is not eligible for the spectral study in a large wavelength range).
In this case, the absorption coefficient $A$ is simplified as:
\begin{equation}
    \label{eq:A}
    A=-log_{10}(T)/d
\end{equation}
where the reflection is not taken into account as the spectral feature is more of interest than the absolute value of absorption.

The birefringence was measured with a polarimeter (Ilis StrainMatic M4/90.50 Zoom) following the S\'enarmont method~\cite{katte2009measuring}.
The phase difference between the two axes with different refractive indices was measured, i.e. the optical retardation $\Gamma$. The birefringence $\Delta n$ is then given by:
\begin{equation}
    \label{eq:Bifi}
    \Delta n = \Gamma/d
\end{equation}
The measurement of the retardation $\Gamma$ was spatially resolved, which gives a map of birefringence $\Delta n$ for the sample.

Single substitutional nitrogen P1 centers were measured to investigate the link between nitrogen doping and diamond absorption/birefringence.
We used two methods to measure their concentration in this work: electron paramagnetic resonance (EPR) and UV-Vis spectroscopy.
The EPR measurement was conducted at room temperature with an EPR spectrometer (Bruker ELEXSYS E580), which is fitted with a Bruker super-high-Q probehead (ER4122 SHQE).
The microwave frequency was set to 9.84~GHz, and the P1 concentration was determined using the built-in spin-counting feature, from the acquisition software (xEPR).
UV-Vis measurements follow our methods paper~\cite{luo2022rapid}. 
We extracted the absorption band at 270~nm from the other spectral features, determine the absorption coefficient plots via Equation~\ref{eq:A} and the concentration of P1 centers via the absorption cross section determined from multiple reference samples.

\section{Diamond absorption}
\subsection{Absorption for varying nitrogen contents}
\label{Sec:abs_N}
\begin{figure}[!htb]
\centering
\includegraphics[width=0.5\textwidth]{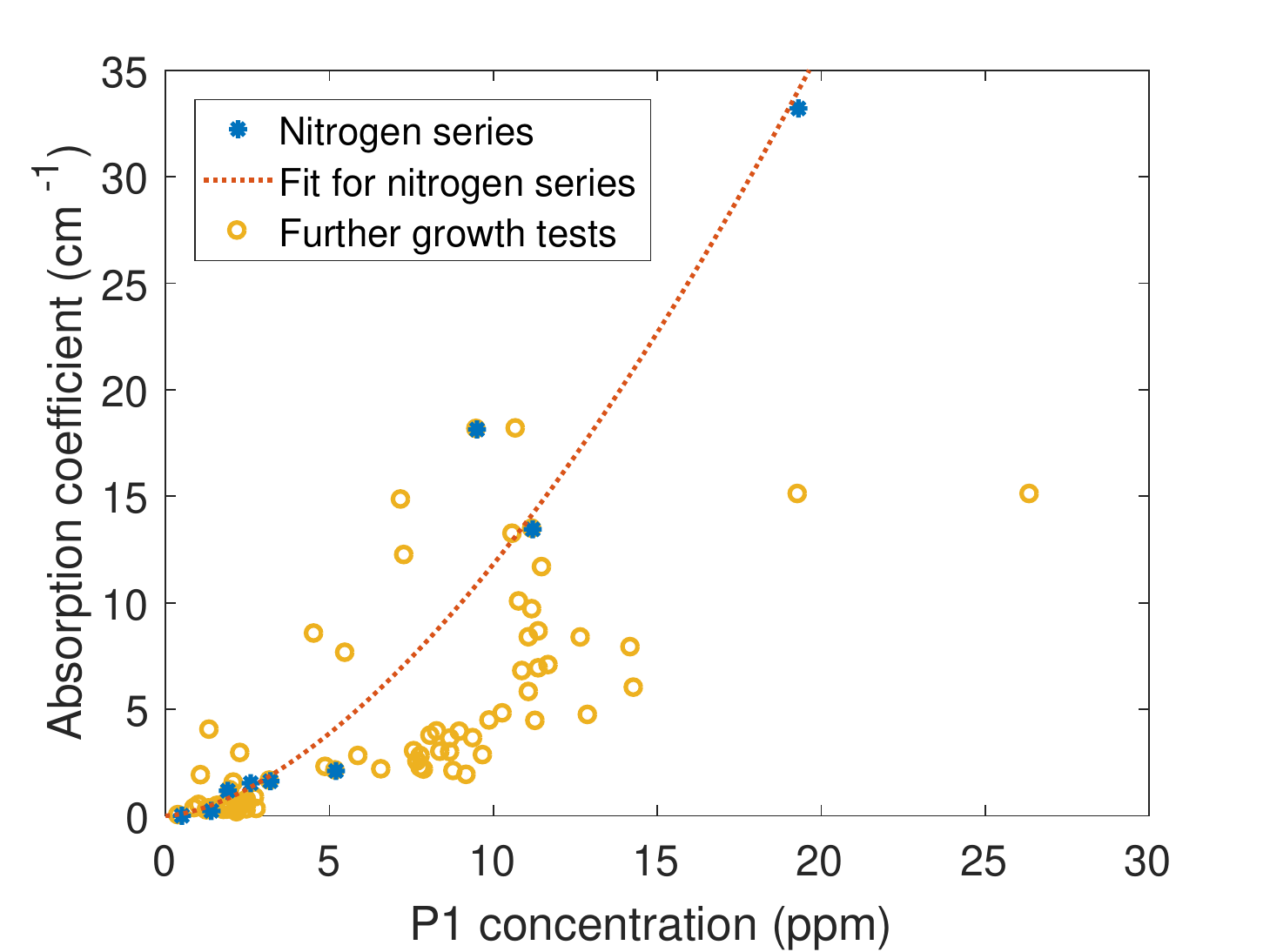}
\caption{Absorption coefficient (Equation~\ref{eq:Acoeff}) for different P1 concentrations in diamond. When growing CVD diamonds with only N/C ratio varying (blue asterisk, `Nitrogen series'), their absorption shows a super-linear correlation (orange dashed line) with the concentration of P1 centers. Further growth tests (yellow circle) in the same reactor confirm the positive correlation between absorption and the nitrogen content in diamonds. This correlation can be optimized by locally optimizing the growth parameters for different nitrogen doping levels.}
\label{fig:Abs_P1}
\end{figure}

To investigate the link between nitrogen content and diamond absorption, we have grown a nitrogen series by varying the N/C ratio (altered by different values of N$_2$ flow for a fixed CH$_4$ flow into the plasma) and keeping other conditions fixed.
For this series, the absorption coefficient shows a super-linear correlation with the as-grown P1 concentration, Figure~\ref{fig:Abs_P1}, blue asterisk for the series and orange dashed line for its fit.
This clearly indicates that nitrogen doping during the CVD growth attributes to diamond absorption, a higher nitrogen doping level leads to stronger absorption.
However, whether the P1 center is a direct contributor to absorption needs further investigation, which will be discussed in the next section.

This preliminary correlation between P1 centers and absorption can be improved (i.e. to reduce absorption for the same P1 concentration) by optimizing the growth parameters.
For that, we have further grown samples in the same reactor with varied individual growth parameters, such as oxygen or methane flow, total gas flow, pressure, and holder geometry, which have been discussed in~\cite{langer2022quantum, langer2022manipulation}.
Yellow circles in Figure~\ref{fig:Abs_P1} show the result of these growth protocols. 
%where the orange dashed line shows the original P1-absorption fit as the same as in Figure~\ref{fig:Abs_P1}(a).
By adjusting growth parameters for different P1 concentrations, we can remarkably reduce the absorption compared to the original fit (orange dashed line).
Since the original recipe was optimised at lower nitrogen doping level the potential for improvement at higher nitrogen doping shows that different growth `recipes' are needed for respective P1 concentrations to achieve an optimized crystal quality. 
Above all, the large number of samples confirms that absorption is positively correlated to the nitrogen content in the diamond, calling for concrete spectral studies for further insight.

\subsection{UV-Vis spectral study}
\begin{figure}[!htb]
\centering
\includegraphics[width=0.5\textwidth]{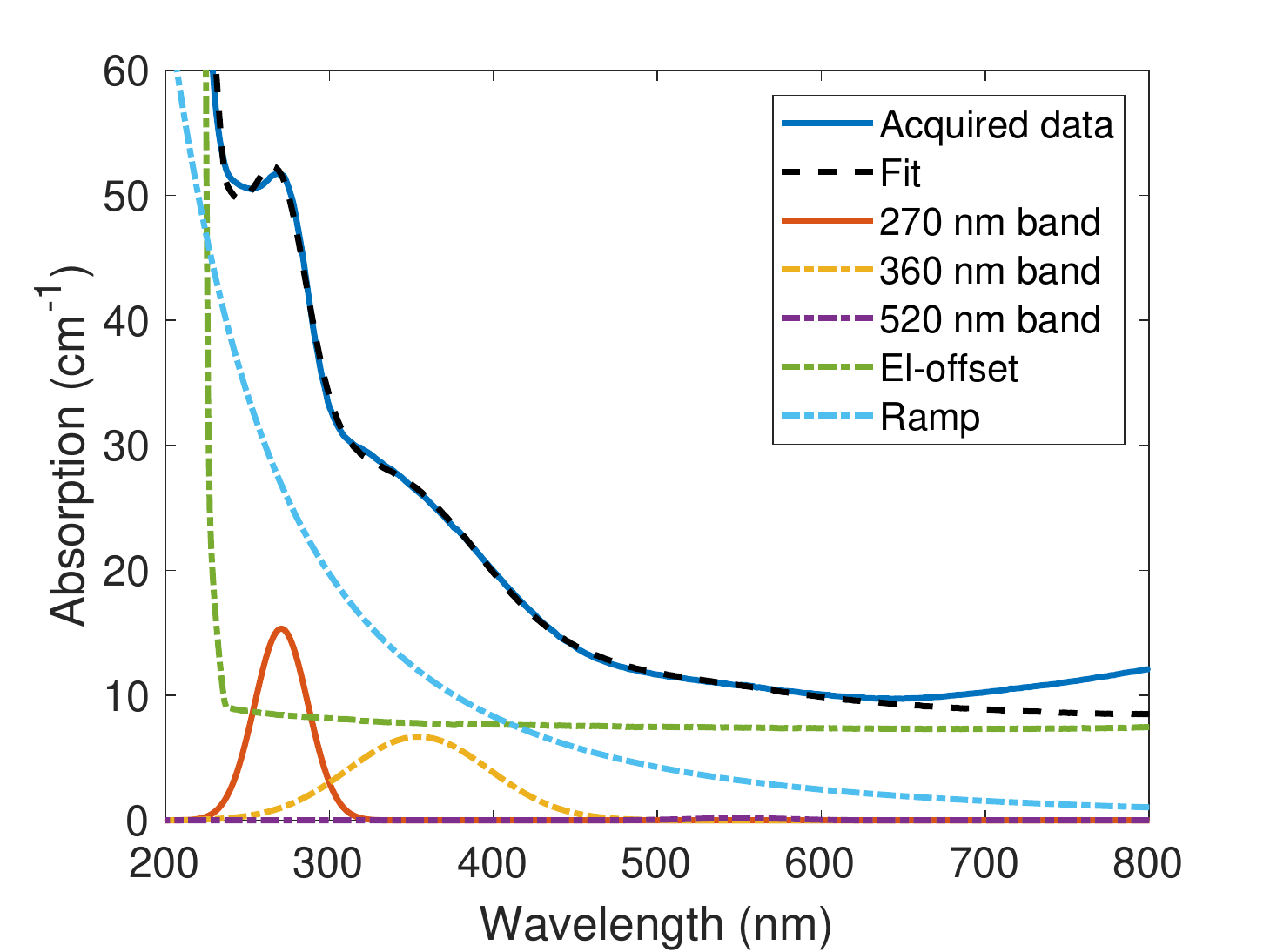}
\caption{The UV-Vis absorption spectrum and the fit of the nitrogen-doped diamond after growth. The offset in the entire UV-Vis range ('El-offset'), the `ramp', and a continuous increase starting from 650~nm are the main contributor to the absorption at $\sim$700~nm.}
\label{fig:Abs_sp}
\end{figure}

To understand the causes of the absorption at 700~nm, we fitted the diamond UV-Vis spectrum and separated it into components to see which are the influencing factors.
We used the method introduced earlier in~\cite{luo2022rapid} for the fitting (Figure~\ref{fig:Abs_sp}): the spectrum is separated into five components, including three Gaussian bands (which are centered at 270~nm, 360~nm, and 520~nm), a `ramp' describing the overall decreasing trend of the spectrum, and a reference spectrum from a pure diamond (`El-offset').
Be aware that all samples in Figures~\ref{fig:Abs_P1}~and~\ref{fig:Abs_sp} are as-grown samples with low NV concentrations, therefore their NV absorption band (approximately at 400-700~nm) is very
weak and has negligible influences.

The 270~nm absorption band links to the P1 center, which has a small bandwidth that shows no effect on the absorption around 700~nm. 
This means, although absorption at 700~nm increases with the P1 concentration, the P1 center is not a direct contributor to the absorption at this regime.
The 360~nm and 520~nm absorption band have been suggested to be related to vacancy clusters and NVH$^0$ centers respectively~\cite{khan2009charge}. 
Both absorption bands do not reach into the relevant wavelength regime around 700~nm and thus do not contribute to NV signal light absorption. 

The spectral fitting indicates three main factors that are accountable for the high absorption: the offset of the spectrum in the entire UV-Vis range (i.e. the fitting coefficient for the `El-offset'), the `ramp', and a continuously increasing feature starting from 650~nm.  
This spectrum is quite representative since these three spectral components are typically the main contributors to the absorption at 700~nm in all samples investigated. The continuously increasing feature starting from 650~nm only appears for the highly nitrogen-doped samples. We now discuss the three features and their potential causes respectively.

The offset of the entire spectrum is hardly assigned to any single defect but more complex causes link to the diamond crystal quality. 
One hypothesis points to non-diamond carbon inclusions, which leads to grayish coloration in the diamond and a high absorption in the whole UV-Vis range~\cite{zaitsev2020nitrogen}.
Our highly nitrogen-doped samples show deep gray color that conforms to this hypothesis, though further verification with Raman spectroscopy is needed.
If this assumption is true, the high absorption for increasing nitrogen content is more likely a fundamental issue in the nitrogen-doped CVD growth.
Nevertheless, it also means that the appearance of the large spectral offset can be potentially reduced by a high-temperature treatment, especially the HPHT annealing, which can transform carbon inclusions back into diamond. 

The `ramp' shows less effect than the offset, it is generally suggested to be vacancy-related~\cite{maki2007effects,jones2009dislocations,luo2022creation}.
% Irradiation and annealing treatments will change the `ramp'~\cite{luo2022creation}, however, the changes at $\sim$700~nm is very limited.
The absorption increase in 650-800~nm is significant for samples with high P1 concentrations.
This indicates the H2 center (NVN$^-$) as a possible candidate: 
the H2 center is formed from the P1 center under similar
conditions with the NV center, meaning a positive correlation between the
as-grown concentration of H2 centers and P1 centers can be expected.
Since the spectrum was only measured up to 800~nm, whether the H2 center is the actual cause needs to be further investigated and could be confirmed by extending the absorption spectrum past its ZPL (at 986.3~nm).

From the large number of the growth tests (Figure~\ref{fig:Abs_P1}), we found that optimizing the growth can reduce the offset of the absorption, and the diamond exhibits less grayish coloration.
The `ramp', however, did not show a clear trend with either the nitrogen doping level or other growth parameters.
The increasing feature from 650 to 800~nm could be hardly removed by optimizing the growth protocol. This feature seems to always occur for the very high nitrogen-doping regime.
In this sense, the optimization of absorption in the growth phase should be mainly focused on reducing the offset.
Further treatments, e.g. high temperature annealing can potentially help to reduce the absorption after growth, which will be our future study.

\subsection{Influence of irradiation and annealing on absorption}
Most of the NV-ensemble-based sensing applications require a high NV concentration, which is hardly achievable in as-grown diamonds.
To create more NV centers in the diamond, irradiation and annealing steps are often conducted. 
Especially electron-beam (e-beam) irradiation is well established for creating single vacancies in bulk diamond plate, a subsequent annealing step then combines vacancies with P1 centers to form NV centers.
Although diamond absorption can be optimized during and after growth, the irradiation and annealing treatment are often the last steps that decide the final state of the crystal quality.
Therefore, it is necessary to understand their influence on absorption.

\begin{figure}[!htb]
    \centering
    \begin{subfigure}[h]{0.49\textwidth}
        \centering
        \includegraphics[width=\textwidth]{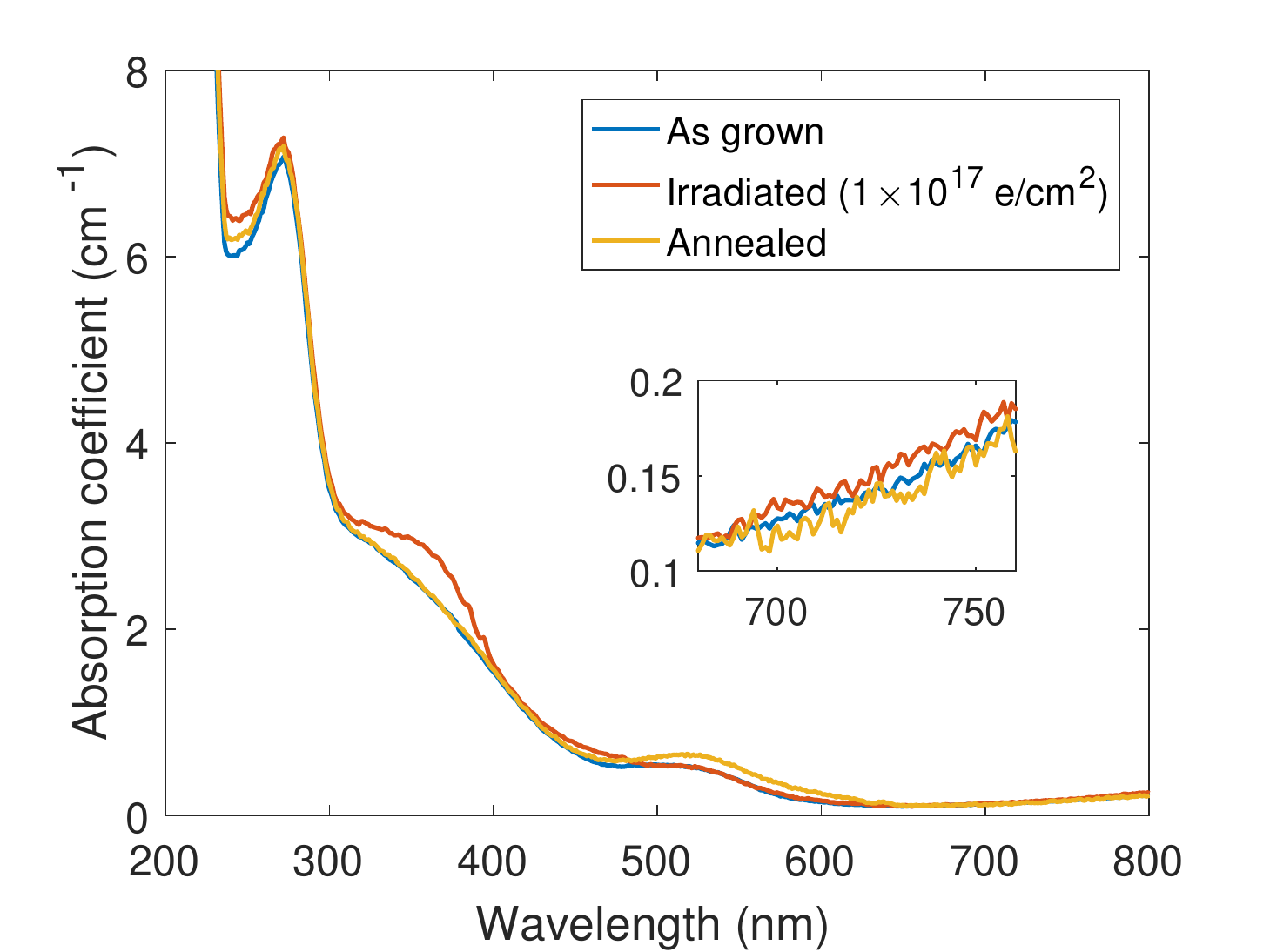}
        \put(-190,135){(a)}
    \end{subfigure}
    \begin{subfigure}[h]{0.49\textwidth}
        \centering
        \includegraphics[width=\textwidth]{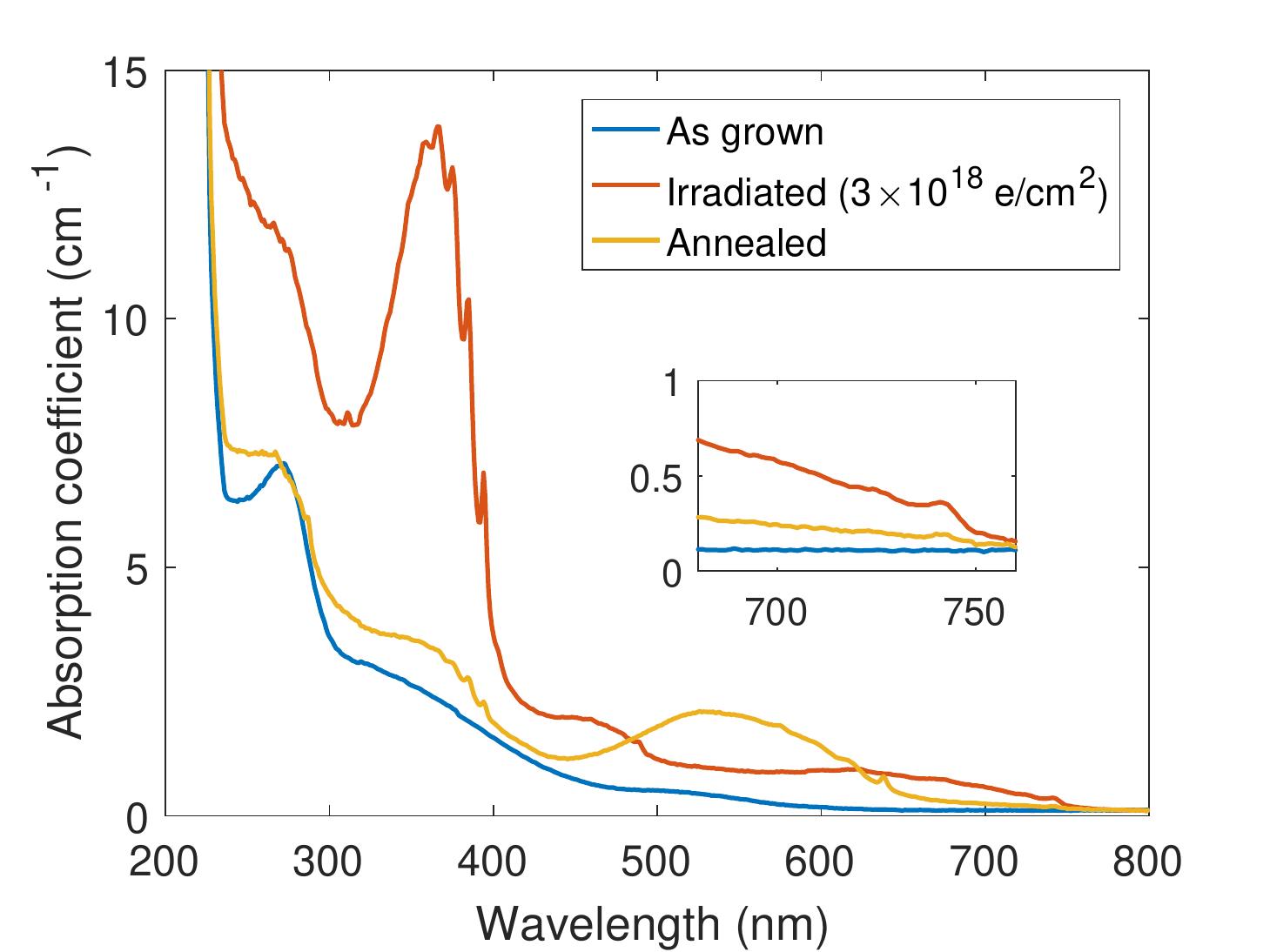}
        \put(-190,135){(b)}
    \end{subfigure}
    \caption{Absorption spectra respectively after growth, irradiation, and annealing steps for (a) 1$\times$10$^{17}$~e/cm$^2$ and (b) 3$\times$10$^{18}$~e/cm$^2$ irradiation fluence (both with 1~MeV electron energy). The annealing step was conducted with 1000$^{\circ}$C for 2~h.}
    \label{fig:Abs_irr}
\end{figure}

We here compare two samples irradiated with 1~MeV electron energy, respectively with 1$\times$10$^{17}$~e/cm$^2$ (as `low fluence') and 3$\times$10$^{18}$~e/cm$^2$ (as `high fluence').
The two samples have been grown under the same conditions, with an initial P1 concentration of 2.2~ppm, and they were annealed after irradiation simultaneously.
We took their UV-Vis spectrum after each step (after growth/irradiation/annealing), Figure~\ref{fig:Abs_irr}, the absorption coefficient was calculated according to Equation~\ref{eq:A}.

For the low irradiation fluence, both the irradiation and annealing show a negligible effect on the absorption at 700~nm, Figure~\ref{fig:Abs_irr}(a).
The irradiation did not create any defect with an absorption band in this regime, and the newly created NV absorption band after annealing has a `tail' that hardly extends to 700~nm (but ends at around 650~nm). 
The absorption coefficient of interest stays consistent after the treatment.

In contrast, for the high irradiation fluence, Figure~\ref{fig:Abs_irr}(b), the creation of neutral vacancies V$^0$ (GR1 centers) significantly increased the absorption at 700~nm, as its absorption band covers a broad range from around 500 to 750 nm.
The annealing treatment converted GR1 centers partially into NV$^0$ centers, leading to a decrease in absorption.
However, GR1 centers were not fully converted, their remaining part still results in a higher absorption than the as-grown phase.
Earlier on we have stated that over-irradiation is harmful~\cite{luo2022creation}, as it creates more NV$^0$ thus deteriorating the NV charge stability and the magnetic resonance contrast.
The result here further supports that statement, as over-irradiation remarkably increases the absorption at 700~nm, which results in a higher optical loss in the material.
The annealing step can compensate for that to some extent (by converting a part of GR1 centers), only in the sense of pulling down the absorption at 700~nm, but still at the cost of promoting the creation of NV$^0$. 
Consequently, one should avoid over-irradiating the diamond, and an optimized irradiation condition is of great importance to acquire high-quality diamonds with high NV concentrations.

The two samples in Figure~\ref{fig:Abs_irr} are representative of multiple CVD samples that we have studied~\cite{luo2022thesis_abs}.
Nevertheless, the influence of the after-growth treatment is in general much less than
the synthesis process. 
Different nitrogen doping levels can cause orders of magnitude increases in absorption (Figure~\ref{fig:Abs_P1}), which is way higher than those caused by the GR1 band. 
The most important role of the treatment is to achieve an improved combination of
high NV concentrations and low absorption.
An optimized treatment can increase the NV concentration in CVD diamonds by more than an order of magnitude~\cite{luo2022creation}, while the absorption at 700~nm remains unchanged.
In this sense, the strategy to achieve good combinations of high NV concentration and low absorption is to engineer absorption during the growth, then irradiate the diamond with optimized conditions to enhance the NV concentration.

\section{Diamond birefringence}
\subsection{Birefringence for varying nitrogen contents}
Diamond is not an inherently birefringent material from its crystal structure. Birefringence is introduced by strain fields that are generated by extended defects and applied stress.
It is highly relevant for applications involving cavities, lasers and polarization.
Here we look into how nitrogen doping affects birefringence: 
as doping introduces a different size of the atom, an increased strain could be expected. 

\begin{figure}[!htb]
\centering
\includegraphics[width=0.85\textwidth]{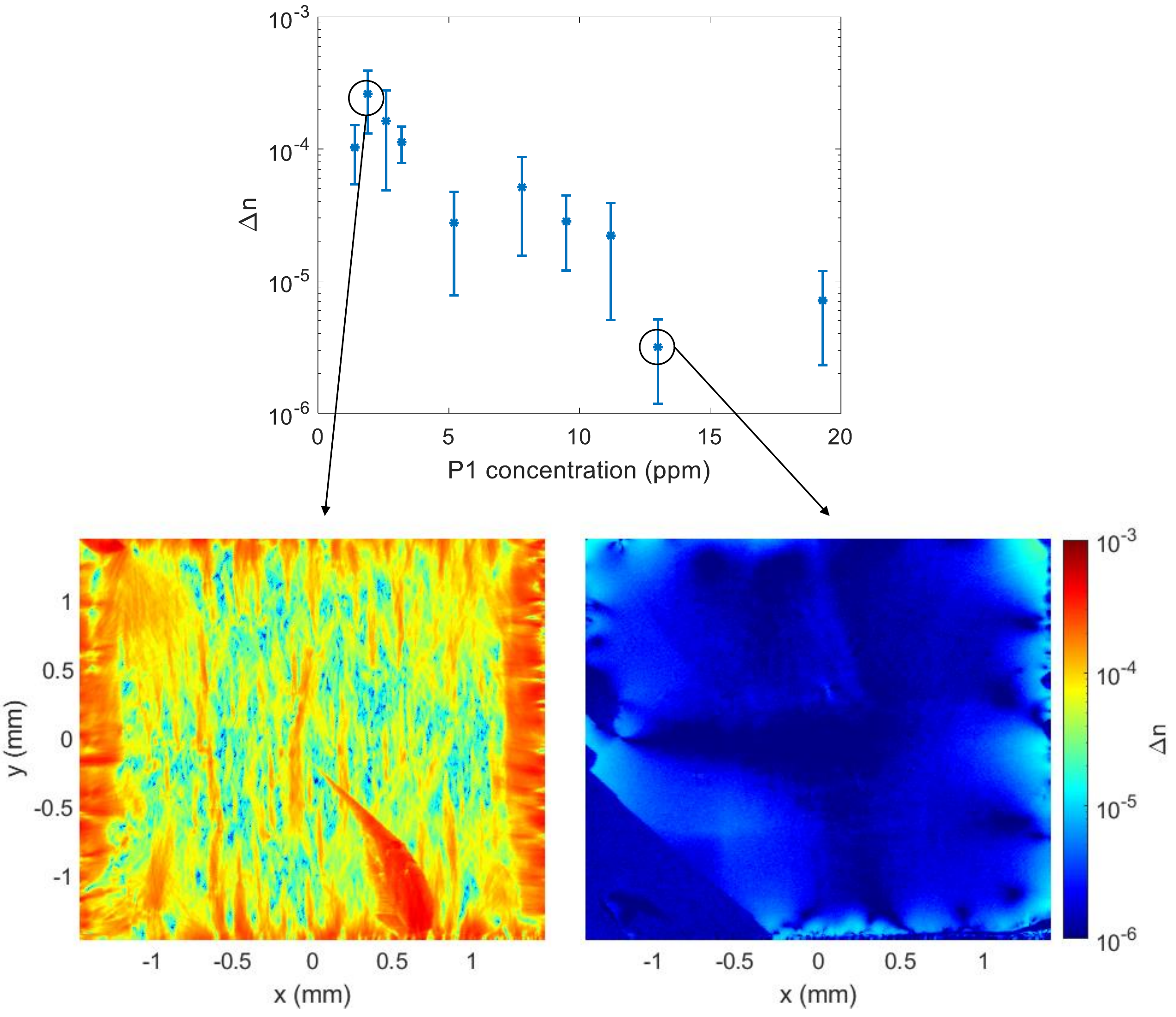}
\put(-270,275){(a)}
\put(-330,135){(b)}
\caption{(a) Birefringence $\Delta n$ decreases for increasing P1 concentrations. The error is given by the standard deviation of the birefringence map (the asymmetry is only due to the logarithmic scaling). (b) The birefringence map of the sample with highest (left) and lowest (right) average $\Delta n$ in (a), which correspond to the second lowest and the second highest P1 concentration respectively.}
\label{fig:Bifi_N}
\end{figure}

We measured the birefringence $\Delta n$ of the nitrogen series (as discussed in Figure~\ref{fig:Abs_P1}) after growth, and plot their average values as a function of the P1 concentration in Figure~\ref{fig:Bifi_N}(a).
The error of birefringence is given by the standard deviation of the spatial variation in the birefringence map of the respective sample. %, the asymmetry is due to the logarithmic scaling in the y-axis.
Both the mean value and standard deviation of the birefringence play roles for the application, as the mean value indicates the total birefringence that an incident beam suffers from, while the standard deviation shows the birefringence inhomogeneity in the sample: 
an ideal material should therefore exhibits a small mean value and a small standard deviation.
Contrary to the absorption that shows a positive correlation with the P1 concentration, birefringence shows a remarkable decrease for an increasing P1 concentration.
We also observed a higher homogeneity of the birefringence when increasing the P1 concentration, as can be seen in error bars of Figure~\ref{fig:Bifi_N}(a). Note that due to the logarithmic scaling on the y-axis, a roughly constant size of error bars to the eye with decreasing mean values on the y-axis means a significant reduction of the absolute value of the errors.
The two samples with the highest P1 concentrations show an average $\Delta n$ below 10$^{-5}$, which has been defined as the standard of `ultra low' birefringence for diamond material in previous work~\cite{friel2009control}.
The theory behind this decreasing trend needs further investigation, but we suggest that either nitrogen atoms can help to prevent the formation of extended defects, or they help to release local stress caused by other defects in the growth.
As a consequence, this negative correlation is a positive sign for the aim of creating highly NV-doped diamonds with low birefringence. 
Figure~\ref{fig:Bifi_N}(b) shows respectively the birefringence map for a low (left) and high (right) P1 concentration as an example. 
Both lower values of birefringence $\Delta n$ and improved homogeneity are clearly visible for the higher nitrogen doping (right), indicating better quality.

Birefringence changes the polarisation of a light beam, which can lead to optical losses or reduced precision in setups that depend on maintaining a polarization, such as a cavity with elements in Brewster's angle or polarisation-selective excitation of specific NV orientations. We can estimate the worst-case-scenario optical loss by assuming the worst relative orientation between the birefringent axis and polarisation direction and assuming that the light shifted out of linear polarisation in a single pass is entirely lost. Then the absolute value of the measured birefringence at $\Delta n= 10^{-4}$ to 10$^{-5}$ translates to a single-pass light-intensity loss between $0$ and a maximum of $\sin^2( \pi \,\Delta n \,d \, / \, \lambda ) = 1.8\%$ to $0.018\%$ respectively through a diamond thickness of $d=300~\mathrm{\mu}$m for a wavelength $\lambda=700$nm. This is typically small compared to absorption losses in highly NV-doped diamond \cite{hahl2022magnetic}.

\subsection{Influence of irradiation and annealing on birefringence}
\begin{figure}[!htb]
\centering
\includegraphics[width=0.9\textwidth]{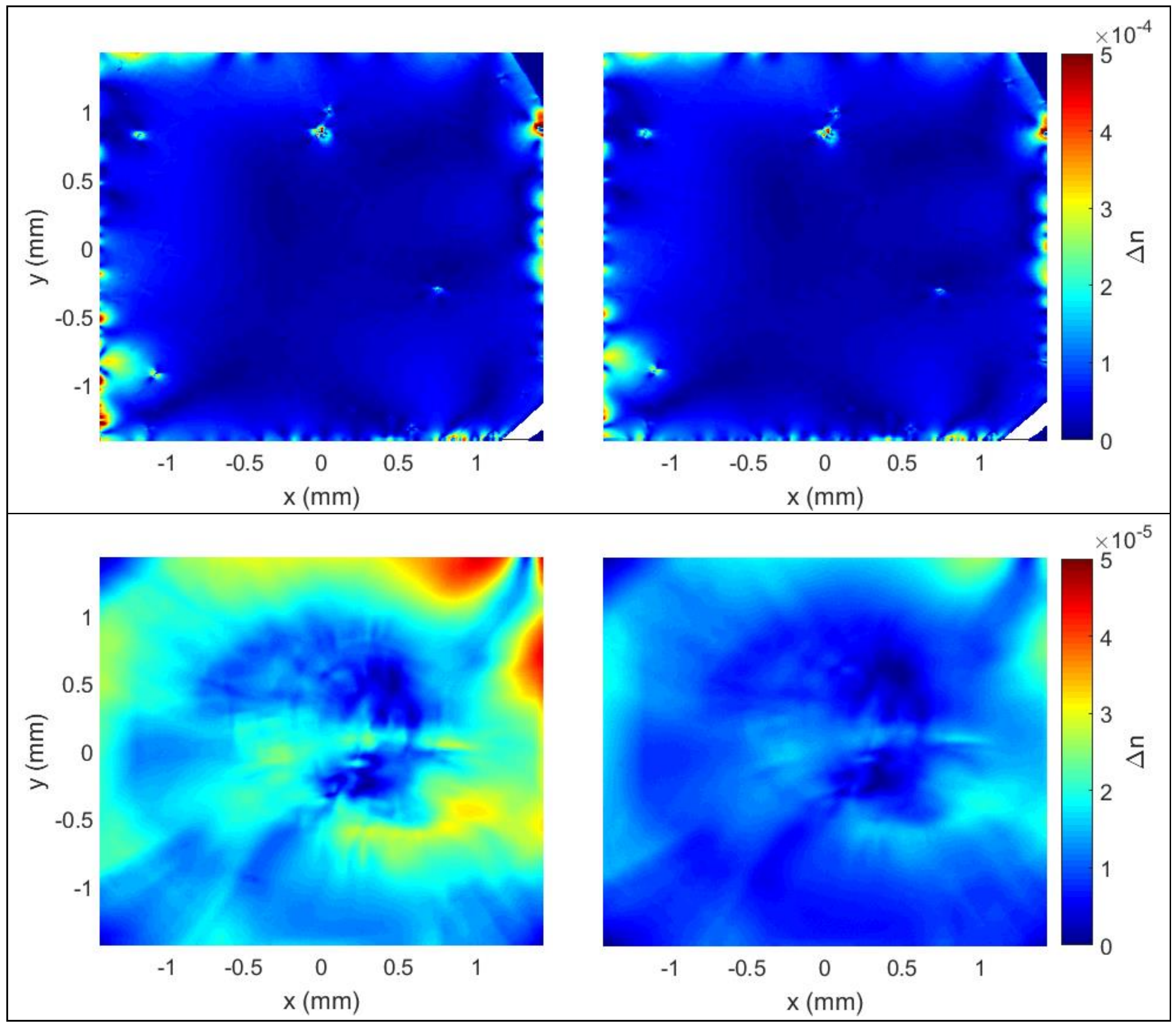}
\put(-340,290){(a)}
\put(-340,140){(b)}
\caption{Birefringence before (left) and after (right) irradiation and annealing. (a) For thin diamond plates (below 500$\mu$m) no significant change was observed. (b) For thick diamond plates (above 1000$\mu$m) a reduction in birefringence over the entire sample area was observed.}
\label{fig:Bifi_irr}
\end{figure}

We also investigated potential birefringence changes by e-beam irradiation and annealing.
The samples were irradiated with 2~MeV electron energy (with different fluences) and then annealed.
We compared birefringence before and after treatments in both CVD and HPHT Ib diamonds, Figure~\ref{fig:Bifi_irr} is representative of the multiple samples that we have investigated~\cite{luo2022thesis_bifi}.
For diamond plates with thickness below 500~$\mu$m (both CVD and HPHT diamonds), no significant change in birefringence has been observed before and after the treatment, Figure~\ref{fig:Bifi_irr}(a).
However, in some thick HPHT diamond plates (>1000~$\mu$m), reduced birefringence has been observed in the whole sample area, Figure~\ref{fig:Bifi_irr}(b). 
The reduction in these thick samples was mainly introduced by the annealing step.
It should not originate from the surface changes, but from the changes in the entire depth of the sample, as the thick samples show a more significant effect in the whole area than the thin samples.
A possible explanation is that vacancies and their positioning in energetically favorable positions (such as NV centers etc.) lead to a small release of local stress, which is similar to the assumption that more nitrogen-doping releases stress in the crystal (Figure~\ref{fig:Abs_P1}).
To conclude, the irradiation and annealing parameters that we used show a minor and positive (i.e.~reducing) influence on diamond birefringence, which benefits the creation of highly NV-doped diamonds with low optical losses.

\section{Conclusion}
In this work, we study diamond absorption and birefringence, with the particular aim to achieve low values for both in diamonds with a high concentration of NV centers. 
This improves diamond quality for sensing applications in general by reducing optical losses and is essential for diamond lasing, LTM, diamond cavity applications, high-power applications, and polarization-dependent applications of NV diamonds. 

We found that the absorption at around 700~nm increases remarkably for increasing P1 concentrations, however, the P1 center itself is not the direct influence factor.
The high nitrogen doping level in the growth leads to high absorption in the whole UV-Vis range (as an `offset') and exhibits a continuous increase from 650-800~nm, which are the main contributors to the high absorption at the regime of interest.
We speculated their causes to be respectively non-diamond inclusions and H2 centers (NVN$^-$), which need further investigation.
Optimizing the growth conditions for individual nitrogen concentrations helps to reduce the absorption within the growth process.
When treating the diamond for an enhanced NV concentration, one should avoid `over-irradiating' the diamond with too high fluences, as it creates neutral vacancies (V$^0$, GR1 band) instead of negatively charged vacancies. These neutral vacancies contribute a large increase of the absorption at 700~nm.
This coincides with the requirement for high NV$^-$/NV ratios, as both the neutral vacancies and neutral NV$^0$ state should be avoided, calling for an optimal irradiation fluence as determined in~\cite{luo2022creation}. If this condition is met an increase of absorption can be avoided in the process of irradiation and annealing as no other influences on the absorption have been measured by the two processing steps.

Opposite to the absorption, the average birefringence of the diamond decreases for increasing P1 concentrations, and samples with high P1 concentrations show better homogeneity of the birefringence: we assume that nitrogen atoms help to release the strain during the growth.
Electron-beam irradiation and subsequent annealing steps show a minor and if any a positive (i.e.~reducing) influence on diamond birefringence.

Overall diamond birefringence is translated to a mostly negligible loss rate, while absorption is a larger factor of optical losses in NV-doped diamond.
A compromise between the demand for high P1 densities (and correspondingly high NV densities) and low absorption must be found for cavity applications.
The optimization of the absorption and birefringence should be mainly considered during the growth.
High-temperature treatments as well as other influences in the growth can potentially reduce the diamond absorption at 700~nm, which will be the direction of our future study.

\section{Acknowledgement}
We thank Alexander Zaitsev, Marco Capelli, Brant Gibson, Andrew Greentree, Brett Johnson, Peter Knittel, Christoph Schreyvogel, Oliver Ambacher and Philippe Bergonzo for valuable discussions. We also thank Michael Ardner, Christine Lell, and Michaela Fritz for preparing the diamond plates; Dorothee Luick for the technical support of UV-Vis measurements; Shangjing Liu for the technical support of birefringence measurements. T.L. and J.J. acknowledge the funding by the German Federal Ministry for Education and Research Bundesministerium für Bildung und Forschung (BMBF) under Grant No. 13XP5063.

\bibliographystyle{unsrt}
\bibliography{references}

\clearpage

\end{document}